\documentclass[twocolumn,showpacs,superscriptaddress,preprintnumbers,amsmath,amssymb]{revtex4}
\usepackage{graphicx}% Include figure files
\usepackage{dcolumn}% Align table columns on decimal point
\usepackage{bm}% bold math
\usepackage{amsmath,amssymb}

\begin{document}

\title{Hydrodynamic collective effects of active proteins in biological membranes}

\author{Yuki Koyano}
\affiliation{Department of Physics, Graduate School of Science, Chiba University, Chiba 263-8522, Japan}

\author{Hiroyuki Kitahata}
\affiliation{Department of Physics, Graduate School of Science, Chiba University, Chiba 263-8522, Japan}

\author{Alexander S. Mikhailov\footnote{Corresponding author. E-mail: mikhailov@fhi-berlin.mpg.de.}}
\affiliation{Abteilung Physikalische Chemie, Fritz-Haber-Institut der Max-Planck-Gesellschaft, 14195 Berlin, Germany}
\affiliation{Department of Mathematical and Life Sciences, Hiroshima University, Hiroshima 739-8526, Japan}

\begin{abstract}
Lipid bilayers forming biological membranes are known to behave as viscous 2D fluids on submicrometer scales; usually they contain a large number of active protein inclusions. Recently, it has been shown [Proc. Nat. Acad. Sci. USA {\bf 112}, E3639 (2015)] that such active proteins should induce non-thermal fluctuating lipid flows leading to diffusion enhancement and chemotaxis-like drift for passive inclusions in biomembranes. Here, a detailed analytical and numerical investigation of such effects is performed. The attention is focused on the situations when proteins are concentrated within lipid rafts. We demonstrate that passive particles tend to become attracted by active rafts and are accumulated inside them.
\end{abstract}


\maketitle

\section{Introduction}

Formed by lipid bilayers, biological membranes possess both elastic and fluid properties. With respect to stretching or bending, they behave as~elastic surfaces and are thus described by the Helfrich model~\cite{Helfrich}. On the other hand, they behave as viscous two-dimensional (2D) lipid fluids within the membrane. Because the viscosity of lipid bilayers is about 1000 times higher than that of water, tangential viscous coupling between the bilayer and the solvent on both sides of it is not effective on the scales shorter than the Saffmann-Delbr{\"u}ck length of about a micrometer~\cite{Delbruck}. On such relatively short length scales, flows in biological membranes are described by the Navier-Stokes equations of classical 2D hydrodynamics (see, e.g., Ref.~\cite{Diamant}). The 2D hydrodynamic effects in biomembranes could be demonstrated by measurement of~diffusion coefficients for small membrane inclusions~\cite{Gambin}. In hybrid numerical simulations, combining simplified molecular dynamics (MD) for lipids with the multiparticle collision dynamics for the solvent, the characteristic 2D dependences of transverse and longitudinal lipid velocity-velocity correlation functions could be seen~\cite{MJ}. Remarkably, there is no linear coupling between shape deformations and lipid flows in biomembranes~\cite{Levine} and, therefore, lipid hydrodynamic and elastic effects can be separately considered.

Typically, membranes are multi-component, i.e., they represent mixtures of different lipid molecules. Due to potential interactions between the lipids, phase separation can take place in biomembranes, leading to the formation of domains enriched with one of the components. According to the classical theory of phase separation, macroscopic phase domains should then be established at long times. Such large domains could indeed be observed in the experiments with artificially created membranes~\cite{Baumgart}. However, macroscopic phase separation is not seen in biological membranes under physiological conditions, and there is much indirect evidence suggesting that small domains, known as lipid rafts, are instead present inside them~\cite{Simons, Hancock}. \textit{In vivo} experiments using stimulated emission depletion (STED) microscopy have narrowed the size range of the rafts to 10-20 nanometers~\cite{Eggeling}. It has been suggested that equilibrium nanoscale rafts can result from micro-phase separation effects that are observed when interactions between the two layers in a membrane are taken into account (see, e.g., Refs.~\cite{Schick, Komura, Schmid, Reigada}).

Biological membranes also contain many protein inclusions, with proteins making up to 40 percent of the total membrane mass. Because of the interactions between them and surrounding lipids, proteins often tend to accumulate within the rafts. Most of membrane inclusions, such as, ion channels, pumps or receptors, are~active and cyclically operate as non-equilibrium protein machines. In each turnover cycle, the protein shape is changed and ligand-induced mechanochemical conformational motions are observed. Since local~coupling between the inclusion and the membrane depends on the protein conformation,~active conformational changes induce non-equilibrium fluctuations of the membrane shape~\cite{Prost, Manneville} that could be experimentally observed~\cite{Girard}. Moreover, interactions between active proteins, mediated by the membrane, become also modified and, as a result, non-equilibrium periodic Turing-like stationary or traveling structures can develop on nanoscales~\cite{Chen}.

Not only membrane shape perturbations, but also hydrodynamic flows can be induced by conformational changes. Because the size of a protein inclusion is larger than the bilayer thickness, parts of a protein protrude into the surrounding solvent. Hence, when active conformational changes occur, hydrodynamical flows in the solvent become generated. The effects of such solvent flows have been taken into account in the previous analysis by treating each active protein inclusion as a hydrodynamical force dipole~\cite{Manneville, Chen}. They contribute to non-equilibrium membrane shape fluctuations and modify membrane-mediated interactions between active proteins themselves.

Recently, it has been noted that lipid membrane flows are also generally induced when conformational changes in inclusions take place~\cite{PNAS}. They can be taken into account by considering an active protein as a force dipole, but with respect to the 2D lipid fluid. Persistent stochastic oscillations of 2D force dipoles corresponding to active membrane proteins give rise to non-equilibrium hydrodynamic fluctuations in the membrane. Advection of passive membrane inclusions in such fluctuating lipid flows leads to substantial diffusion enhancement of such particles and, when the distribution of active proteins in the membrane is not uniform, also to the chemotaxis-like drift~\cite{PNAS}. However, only general expressions for the diffusion enhancement and the drift velocity in biological membranes, as well as simple numerical estimates for the magnitudes of the involved effects,~ have been reported so far.

Here, we perform further analysis of collective lipid hydrodynamic effects of active proteins in biological membranes. Numerical simulations for the evolution of the distribution of passive inclusions when active proteins are spatially localized within small rafts are also performed.

Several simplifying assumptions are employed by us. The membrane is modeled as 2D fluid and three-dimensional (3D) effects of coupling to the solvent are not taken into the account. This limits the applicability of our analysis to the submicrometer length range. Since our attention is focused on hydrodynamic membrane effects, effects of active inclusions on the membrane shape are not included into the present analysis; they have been extensively investigated elsewhere before. Moreover, we consider the spatial distribution of active protein inclusions as given and do not discuss how it could have become formed. While actual biomembranes are usually multi-component and this is important for the development of rafts, possible multi-component hydrodynamic effects are also not yet discussed. Moreover, planar orientations of force dipoles corresponding to active proteins are assumed to be random and statistically independent, neglecting a possibility that nematic orientational in-plane order of active protein inclusions takes place.

In the next section, we show how analytical expressions for diffusion enhancement and drift velocity, induced by active membrane proteins, can be cast into a simpler form. We also write down the evolution equations for passive  particles in the membrane and pay special attention to the drift effects. Then we discuss the possibility of the diffusion enhancement in the actual biological membranes. In Section~\ref{sec-circular}, analytical expressions for diffusion enhancement and drift of particles in hydrodynamic fluctuating fields of active proteins confined to a flat circular raft are derived. In Section~\ref{sec-steadystate}, stationary-state distribution of the small passive particles are discussed considering the diffusion enhancement induced by the active proteins.  Results of numerical simulations for the evolution of small particles are reported in Section~\ref{sec-numerical}. The paper ends with conclusions and a brief discussion of further perspectives in this research.

\section{Evolution equation for passive particles\label{sec-enhancement}}

As shown in Ref.~\cite{PNAS}, conformational activity of membrane proteins leads to the development of lipid flows in the biological membrane. Advection in such non-equilibrium fluctuating flows at low Reynolds numbers results in diffusion enhancement and drift of passive particles. The Fokker-Planck equation for the probability density $p(\bm{r},t)$ of passive particles~\cite{FP} is
\begin{align}
\frac{\partial p(\bm{r}, t)}{\partial t} =& -\frac{\partial}{\partial r_{\alpha}} \left [ V_{\alpha}(\bm{r}) p(\bm{r}, t)\right ] \nonumber \\ &+ \frac{\partial^2}{\partial r_{\alpha} \partial r_{\alpha'} } \left [ D_{\alpha \alpha '}(\bm{r}) p(\bm{r}, t) \right ], \label{timeevolution}
\end{align}
with $\alpha =1 ,2$ and $\bm{r} =(r_1 ,r_2)$ and summation over the repeated indices is assumed. $\delta_{\alpha \alpha'}$ is the Kronecker delta, which is 1 for $\alpha = \alpha'$, and 0 for $\alpha \neq \alpha'$. The matrix of diffusion coefficients $D_{\alpha \alpha '}\left (\bm{r} \right ) =D_{T}(\bm{r})\delta _{\alpha \alpha '} +D_{\alpha \alpha '}^{A}(\bm{r})$ where~$D_{T}(\bm{r})$ is the equilibrium diffusion coefficient that may vary within the membrane. The diffusion enhancement is given by~\cite{PNAS}
\begin{align}
D_{\alpha \alpha'}^{A}(\bm{r}) =& \Omega _{\beta \beta' \gamma \gamma'} \nonumber \\ &\times \int d \bm{r}' \frac{\partial G_{\alpha \beta}}{\partial r'_{\gamma }}\frac{ \partial G_{\alpha' \beta'}}{ \partial r'_{\gamma '}}S_{A}(\bm{r} +\bm{r}')c(\bm{r} +\bm{r}'). \label{eq2}
\end{align}

The velocity is~\cite{PNAS}
\begin{align}
V_{\alpha }(\bm{r}) =& -\Omega _{\beta \beta' \gamma \gamma'} \nonumber \\ & \times \int d \bm{r}' \frac{ \partial^{2} G_{\alpha \beta}}{\partial r'_{\gamma} \partial r'_{\delta }}\frac{ \partial G_{\delta \beta'}}{ \partial r'_{\gamma'}}S_{A}(\bm{r} +\bm{r}')c(\bm{r} +\bm{r}'). \label{eq3} 
\end{align}
In Eqs.~\eqref{eq2} and \eqref{eq3}, $G_{\alpha\beta}$ is the mobility tensor; we have $\Omega _{\beta \beta' \gamma \gamma '} =\left (1/8\right )(\delta _{\beta \beta'} \delta _{\gamma \gamma'} +\delta _{\beta \gamma}\delta _{\beta' \gamma'} +\delta _{\beta \gamma'}\delta _{\beta' \gamma})$.

In the Oseen approximation, the 2D mobility tensor is~\cite{Diamant}
\begin{equation}
G_{\alpha \beta }(\bm{r}) =\frac{1}{4\pi \eta} \left [ -\left (1 +\ln \left (\kappa r \right )\right )\delta _{\alpha \beta} +\frac{r_{\alpha}r_{\beta}}{r^{2}} \right ], \label{2DOseen}
\end{equation}
where $r = \left| \bm{r} \right|$, $\eta$ is the 2D viscosity of the membrane, and ${\kappa}^{-1}$ is the characteristic Saffman-Delbr{\"u}ck length, $\kappa^{-1} = \eta h/(2\eta_{s})$, where $h$ is the thickness of the membrane and $\eta_{s}$ is the viscosity of the solvent. The local concentration of active proteins is $c(\bm{r})$ and the degree of their activity is characterized by the intensity of force dipoles $S_{A}(\bm{r}).$ Note that the activity $S_A$ depends on the local concentration of ATP or other substrates needed by active proteins to cycle. The integration is performed over the entire membrane. Equations~\eqref{eq2} and \eqref{eq3} are derived assuming that the 2D force dipoles corresponding to different proteins are statistically non-correlated and that they are randomly oriented in the membrane plane~\cite{PNAS}. Coupling to the solvent is neglected in these equations, limiting the description to membrane regions with the size shorter than the Saffman-Delbr{\"u}ck length.

Because of the four summations, previously derived~\cite{PNAS} general expressions~\eqref{eq2} and \eqref{eq3} are complicated. Now, we could show (the derivations are given in Appendix~\ref{app1}) that, under the 2D Oseen approximation, these results can be cast in to a more simple and transparent, but equivalent form, i.e.
\begin{align}
D_{\alpha \alpha'}^{A}(\bm{r}) =\frac{1}{32\pi^{2}\eta^{2}}\int d\bm{r}' \frac{r'_{\alpha} r'_{\alpha'}}{{r'}^{4}}Q(\bm{r}+ \bm{r}'), \label{eq5}
\end{align}
\begin{equation}V_{\alpha}(\bm{r}) =\frac{1}{32\pi^{2}\eta^{2}}\int d\bm{r}' \frac{r'_{\alpha}}{{r'}^{4}}Q(\bm{r} + \bm{r}'), \label{eq6}
\end{equation}
where the combination $Q(\bm{r}) = S_{A}(\bm{r}) c(\bm{r})$ has been introduced.

The evolution equation for the concentration $n(\bm{r},t)$ of passive particles can also be obtained. Generally, this equation includes both the drift and the diffusion terms
\begin{align}
\frac{ \partial n(\bm{r} ,t)}{\partial t} =& -\frac{\partial}{\partial r_{\alpha}} \left [ U_{\alpha }(\bm{r}) n(\bm{r}, t) \right ] + \frac{ \partial }{ \partial r_{\alpha }} \left[ D_{\alpha \alpha '} (\bm{r}) \frac{ \partial n(\bm{r} ,t)}{ \partial r_{\alpha'}} \right], \label{eq7}
\end{align}
where
\begin{equation}U_{\alpha }(\bm{r}) =V_{\alpha }(\bm{r}) -\frac{ \partial D_{\alpha \alpha '}(\bm{r})}{ \partial r_{\alpha'}}. \label{eq8}
\end{equation}
Substituting Eqs.~\eqref{eq5} and \eqref{eq6} into Eq.~\eqref{eq8} and performing several transformations of the integrals, a simple approximate expression for the drift velocity has been derived (see Appendix~\ref{app1}),
\begin{equation}
U_{\alpha }(\bm{r}) \simeq \frac{1}{32\pi \eta ^{2}}\frac{ \partial Q(\bm{r})}{ \partial r_{\alpha }}. \label{eq9}
\end{equation}

Using newly derived equations~\eqref{eq5}, \eqref{eq6} and \eqref{eq9}, effects of diffusion enhancement and drift of passive particles in biological membranes can be further analyzed.

It can be readily noticed that diffusion enhancement~\eqref{eq5} is nonlocal, i.e., the diffusion is determined not only by the local concentration of active proteins, but also by their distribution in the neighborhood of it. According to Eq.~\eqref{eq5}, the contributions from active proteins are inversely proportional to the square of the distance, leading to the logarithmic divergence of the integral~\eqref{eq5} at large length scales. It should be however recalled that our 2D theory is limited to the membranes of the size shorter than the Saffman-Delbr\"{u}ck length about a micrometer. For larger membranes, the results hold only assuming that active proteins are localized within a submicrometer-size membrane area and absent outside of it. It can be moreover noted that the diffusion integral~\eqref{eq5} diverges logarithmically at short distances and therefore a cut-off needs to be introduced. Generally, diffusion enhancement~\eqref{eq5} is anisotropic and the anisotropy is controlled by the asymmetry in the distribution of active proteins near the observation point. 

The nonlocality and the logarithmic divergence at short distances are also characteristic for the velocity~\eqref{eq6} that enters into the Fokker-Planck equation~\eqref{timeevolution}. However, such effects become cancelled when the chemotaxis-like drift velocity is calculated according to Eq.~\eqref{eq8}. Remarkably, the drift of passive particles in the membranes is always determined by the local gradient of $Q(\bm{r})$.
 
If the activity of proteins is uniform over the membrane (e.g., if ATP or other substrates are uniformly supplied), the drift velocity is proportional to the gradient of protein concentration and passive particles tend to drift to the regions where active proteins are concentrated. If proteins are uniformly distributed, but their activity level described by variable $S_A$ varies over the membrane, drift of passive particles into the higher-activity areas should take place.

While this effect looks similar to chemotaxis, there are important differences as well. Passive particles drift not because there are physical forces acting on them, and the mobility of passive particles or their size does not enter into Eq.~\eqref{eq9}. 
    
Suppose that active proteins are concentrated within some area (a ``raft'') and the distribution of proteins is approximately uniform within such area and outside it. Moreover, let us assume that the activity of all proteins is the same, $S_{A}(\bm{r}) = \mathrm{const.}$, so that $Q(\bm{r}) =S_{A}c(\bm{r})$.~ Since protein concentration~$c$ is varying only at the boundary of the raft and the drift velocity of passive particles in Eq.~\eqref{eq9} is proportional to the concentration gradient, the drift will be present only in the interface area. Moreover, the direction of the drift velocity coincides, according to Eq.~\eqref{eq9}, with the direction of the drift, implying that, within the interface, the particles will tend to move inside the raft. Outside of the raft and also inside it, only diffusive motions takes place. Thus, passive particles will be effectively adsorbed by the raft of active proteins. Whenever a particle enters the raft boundary, it is dragged inside the raft.~The same behavior should be found if the proteins are uniformly distributed, but their activity $S_{A}$ is enhanced in some area. Then, the region with the enhanced activity will tend to accumulate passive particles inside it. The activity of proteins depends on the local concentration of ATP or other substrates and it can be also chemically regulated, enhanced or inhibited.

Will the effects depend on the size of passive particles? As already mentioned, the drift velocity~\eqref{eq9} is independent of their size. There is however a weak size dependence in the diffusion enhancement given by Eq.~\eqref{eq5}. At short distances $r$, the integral in this equation is logarithmically diverging and therefore a cut-off needs to be introduced. As the cut-off length $\ell_{c}$, the sum of the radii of the active protein ($\ell_{p}$) and of the passive particle ($\ell_{0}$) can be chosen in the simplest approximation, so that $\ell_{c} =\ell_{p} +\ell_{0}$.~ With such cut-off,~diffusion enhancement has the logarithmic size dependence, $D_{A} \propto \ln [\kappa (\ell_{p} + \ell_{0})] .$ Hence, for passive particles with sizes smaller or comparable with that of a protein, the size dependence is practically absent. Note that, because the far-field Oseen approximation has been employed, effects for big passive particles, with the sizes much larger than that of the proteins, cannot be considered here.

So far, we have assumed that only one kind of active proteins is present in the membrane. The extension to the multi-component case is however straightforward. If different kinds $i$ of active proteins are present, each with its own concentration $c_{i}$ and activity~$S_{A ,i}$, Eqs.~\eqref{eq5}, \eqref{eq6}, and \eqref{eq9} still hold, but we should use $Q(\bm{r}) =\sum_{i}S_{A ,i}(\bm{r})c_{i}(\bm{r})$ instead.

\section{Diffusion effects of active rafts\label{sec-circular}}

Suppose that the concentration of active proteins is constant, $c =c_{0}$, within a circular region (``raft'') of radius $R$ whose center is located at the origin of coordinates, and that active proteins are absent, $c=0$, outside of such region. Moreover, the activity of proteins is uniform over the membrane, $S_{A}(\bm{r}) = \mathrm{const.} =S_{A}$, so that $Q(r) =S_{A}c_{0}$ for $r \leq R$ and $Q(r) =0$ for $r >R$.

Substituting this expression into Eq.~\eqref{eq5} and taking the integral, we find that inside the raft, i.e. for $r < R - \ell_c$, diffusion enhancement is isotropic and given by
\begin{equation}
D^{A}(\bm{r}) = \pi \xi \ln \left( \frac{\sqrt{R^2-r^2}}{\ell_{c}} \right ), \label{diffin}
\end{equation}
where $\xi = S_A c_0 /(32 \pi^2 \eta^2)$.
It reaches its maximal value, $D_A = \pi \xi \ln(R/\ell_c)$ at the center of the raft. Note that this value depends logarithmically on the cut-off length $\ell_c$.

Outside of the raft (for $r > R + \ell_c$), diffusion enhancement is anisotropic; it is different along the radial and transverse directions with respect to the raft.  The radial component $D_{\| }^A$ is given by
\begin{equation}
D_{\| }^{A} =\pi \xi \left [ \ln \left( \frac{r}{\sqrt{r^2 -R^2}} \right) +\frac{R^2}{2r^2} \right], \label{diffinn}
\end{equation}
and the transverse component $D_{\bot}^A$ is
\begin{equation}
D_{\bot}^{A} = \pi \xi \left [\ln \left( \frac{r}{\sqrt{r^2 -R^2}} \right) -\frac{R^2}{2r^2}\right ]. \label{diffout}
\end{equation}
Note that these expressions do not involve the cut-off length (see Appendix~\ref{app2}). 

\begin{figure}
\begin{center}
\includegraphics{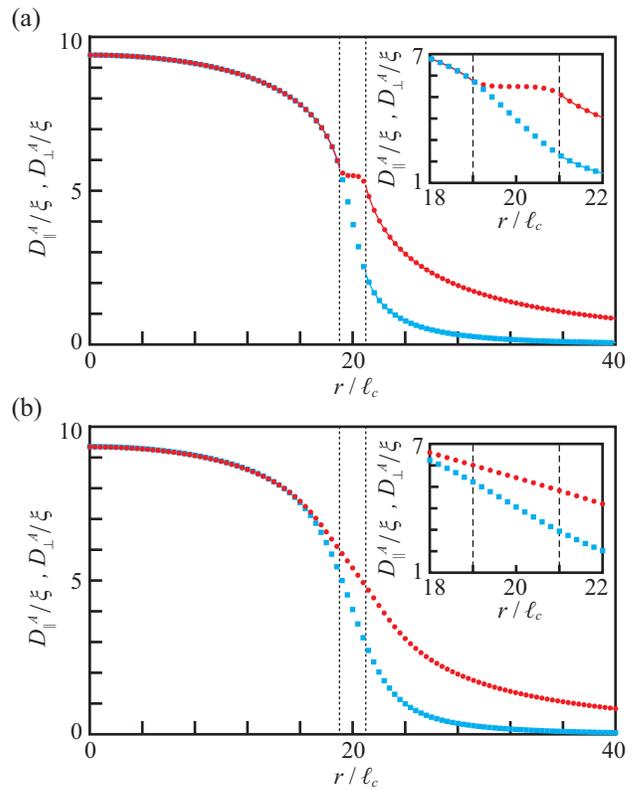}
\end{center}
\caption{(Color online) Diffusion enhancement by a circular raft of radius $R = 20 \ell_c$ with the boundary width (a) $\delta = 0$ and (b) $\delta = 2 \ell_c$. The profiles of diffusion enhancement in the radial, $D^{A}_{\|} / \xi$ (red, dark gray), and in the transverse, $D^{A}_{\bot}/ \xi$ (cyan, light gray), directions as a function of the distance from the raft center are shown.
}
\label{Fig_profile_delta1}
\end{figure}

Finally, inside a narrow ring with $R - \ell_c < r < R + \ell_c$, the integral~\eqref{eq5} cannot be analytically determined,  but it can still be numerically evaluated (see Fig. \ref{Fig_profile_delta1}).  Cross-diffusion is absent due to the symmetry implications.

The asymptotic behavior far from the raft, i.e. at $r \gg R$, can be further considered. In the leading orders of magnitude, we find that
\begin{equation}
D_{\|}^{A} \simeq \pi \xi \frac{R^{2}}{r^{2}}, \label{DAperp}
\end{equation}
\begin{equation}
D_{\bot}^{A} \simeq \frac{\pi \xi}{4} \frac{R^4}{r^4}.
\end{equation}
Hence, the radial component of diffusion falls as the inverse square of the distance to (the center of) the raft, whereas the transverse component depends on the inverse fourth power of this variable and hence it falls much faster than the radial component.

Figure~\ref{Fig_profile_delta1} shows the dependences of the radial and transverse diffusion enhancement on the distance from the center of the raft for sharp ($\delta = 0$, Fig.~\ref{Fig_profile_delta1}a) and smooth ($\delta = 10$, Fig.~\ref{Fig_profile_delta1}b) interfaces. In numerical simulations, the distribution of protein concentration within the raft is given by
\begin{equation}
c(\bm{r}) =\frac{1}{2} c_{0} \left [ 1 + \tanh \left ( -\frac{r -R}{\delta} \right ) \right ], \label{distrib}
\end{equation}
where $\delta$ is the raft boundary width.

The drift velocity $\bm{U}$ is given by Eq.~\eqref{eq9} and can be analytically determined for the distribution in Eq.~\eqref{distrib}. It is always radially directed and pointed towards the center of the raft. The velocity magnitude is given by
\begin{equation}
\left | \bm{U} (\bm{r}) \right | \simeq \frac{\pi \xi}{2\delta} \cosh^{-2} \left ( -\frac{r-R}{\delta} \right ).
\end{equation}
The velocity vanishes outside of the interface, it diverges in the limit~$\delta \rightarrow 0$ of a sharp interface.

Thus, hydrodynamic collective effects of active proteins lead to diffusion enhancement not only inside the raft, but also around it. Outside of the raft, diffusion enhancement is stronger in the radial direction. In contrast to diffusion enhancement, the chemotaxis-like drift velocity~\eqref{eq9} is determined by the local concentration gradient of active proteins. Therefore, the drift is present only at the boundary of the raft and it is directed inwards. 

\begin{figure}
\begin{center}
\includegraphics{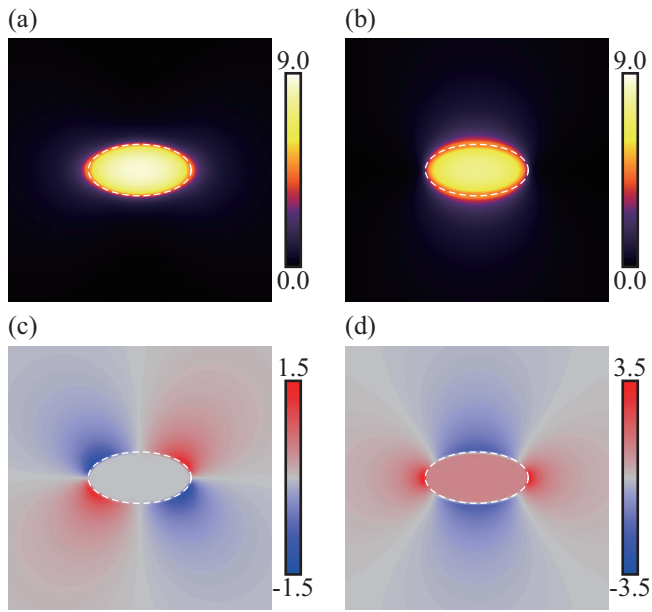}
\end{center}
\caption{(Color online) Diffusion enhancement for an elliptic raft with the semiaxes  $20 \ell_c$ and $10\ell_c$ and the sharp boundary. The diffusion enhancement components (a) $D^A_{11} / \xi$, (b) $D^{A}_{22}/ \xi$, and (c) $D^{A}_{12}/ \xi$ are displayed. The diffusion anisotropy $(D^A_{11} - D^A_{22}) / \xi$ is additionally shown in panel (d).
}
\label{Fig_profile_ellipse}
\end{figure}

We also numerically calculated the profile of the diffusion enhancement in the case of an elliptic raft as shown in Fig.~\ref{Fig_profile_ellipse}. Note that, in contrast to the case with a circular raft, the anisotropy of the diffusion enhancement, i.e., $(D_{11}^A - D_{22}^A)/\xi$, is present also inside the raft.

The orders of magnitude of diffusion enhancement for a typical active raft can be numerically estimated (cf.~\cite{PNAS}). The 2D lipid viscosity $\eta$ is expressed as $\eta = \eta_{3\mathrm{D}} h$, in terms of the membrane thickness $h$ and the 3D lipid viscosity $\eta_{3\mathrm{D}}$. We can take $\eta_{3\mathrm{D}} = 1 \, \mathrm{Pa \cdot s}$ and $h = 1\, \mathrm{nm}$. The force dipole activity of typical proteins has been previously estimated~\cite{PNAS} to be about $S_A = 10^{-43} \, \mathrm{N^2 \cdot m^2 \cdot s}$. We assume that the raft has the radius of $R = 100 \, \mathrm{nm}$ and it contains about 10 active proteins, so that their membrane concentration is of the order of $10^{15} \, \mathrm{m^{-2}}$. Assuming that passive particles are not larger than proteins, the cut-off length is chosen as $\ell_c = 5 \, \mathrm{nm}$. With these numerical values, the maximal diffusion enhancement in the center of the raft is of the order of $D_A = 10^{-8} \, \mathrm{cm^2/s}$.

This numerical estimate should be compared with the equilibrium diffusion constants in a biomembrane. Typically, for proteins in lipid bilayers $D_T = 10^{-10} \, \mathrm{cm^2/s}$ and for lipids $D_T = 10^{-8} \, \mathrm{cm^2/s}$. Hence, according to our estimates, diffusion enhancement inside a raft and in its vicinity should be comparable with the thermal diffusion for small molecules of a nanometer size, and dominate over it for the particles with the size of a protein. We should however stress that the above estimate is very rough, particularly because the value of $S_A$ is not experimentally determined.

\section{Distribution of passive particles\label{sec-steadystate}}

Evolution of the distribution of passive particles in the membrane under the hydrodynamic effects of active proteins is described by Eq.~\eqref{eq7} with the matrix of diffusion coefficients~\eqref{eq5} and the drift velocity~\eqref{eq9}. This equation can be numerically integrated and, moreover, analytical solutions for stationary distributions can also be constructed. 

First, we consider the case when active proteins are localized within a circular raft. 
In the axially symmetric system, the time evolution of the distribution of passive particles, $n(r)$, can be described as
\begin{equation}
\frac{\partial n}{\partial t} =  - \frac{1}{r} \frac{\partial}{\partial r} \left(r U_\| n \right) + \frac{1}{r} \frac{\partial}{\partial r} \left(r D_\| \frac{\partial n}{\partial r} \right),
\end{equation}
and the stationary distribution of $n(r)$ should satisfy
\begin{equation}
- U_{\|}(r) n(r) + (D_T(r) + D_{\|}^A(r)) \frac{\partial n}{\partial r} = 0,
\end{equation}
whose solution is 
\begin{equation}
n(r) = n_{in} \exp \left(\int^r_0 \frac{U_{\|}(r')}{D_T(r') + D_{\|}^A(r')} dr'\right), \label{st_sol}
\end{equation}
where $n_{in}$ is the concentration in the center of the raft. 

If active proteins are uniformly distributed within a raft of radius $R$, so that $c(r) = c_0$ for $r < R$ and $c(r) = 0$ for $r > R$, and their activity level is constant,  $S_A(r) = S_A$, the drift velocity is 
\begin{equation}
U_{\|}(r) = - \pi \xi \delta(r- R). \label{delta_fn}
\end{equation}
Therefore, in this case, passive particles are uniformly distributed inside and outside of the raft in the stationary state, i.e. $n(r) = n_{in}$ for $r < R$ and $n(r) = n_{out}$ for $r > R$, and we have 
\begin{equation}
\frac{n_{in}}{n_{out}} = \exp \left( \frac{\pi\xi}{D_T(R) + D_{\|}^A(R) } \right), \label{nin_nout}
\end{equation}
where $D_T(R)$ is the value of the equilibrium diffusion coefficient at the boundary of the raft. 
As an approximation for diffusion enhancement at the boundary of the raft, we can take $D_{\|}^A(R) \simeq \left(D_{\|}(R - \ell_c) + D_{\|}(R + \ell_c)\right)/2$ and use Eqs.~\eqref{diffin} and \eqref{diffout} to obtain
\begin{equation}
D_{\|}^A(R) \simeq \frac{\pi \xi}{2} \left(\ln \frac{R}{\ell_c} + \frac{1}{2}\right),
\end{equation}
provided $R \gg \ell_c$. Substituting this into Eq.~\eqref{nin_nout}, we find
\begin{equation}
\frac{n_{in}}{n_{out}} = \exp \left( \frac{2}{\ln(R/\ell_c) + (1/2) + (2 D_T /(\pi \xi))} \right).
\end{equation}
If the diffusion enhancement due to active proteins dominates over equilibrium diffusion, i.e., $\xi \gg D_T(R)$, and therefore
\begin{equation}
\frac{n_{in}}{n_{out}} = \exp \left( \frac{2}{\ln(R/\ell_c) + (1/2)} \right), \label{ratio}
\end{equation}
If, for example, $R/\ell_c =20$, this ratio is $n_{in}/n_{out} \simeq 1.77$.

Because their equilibrium diffusion constants are smaller, the large particles like passive proteins are strongly attracted to the circular raft, while small particles such as small molecules are less affected by the active protein raft.

\section{Numerical simulations\label{sec-numerical}}

We have performed numerical simulations of equation~\eqref{timeevolution} that governs evolution of the concentration distribution $n(\bm{r})$ of passive particles. In the simulations, active proteins were present only within the rafts of circular or elliptic shapes. The initial distribution of passive particles was uniform. The model parameters were chosen in such a way that diffusion enhancement was of the same order as thermal diffusion, i.e. $\xi/D_T = 1.$ No-flux boundary conditions were used; the linear size of the system was $L = 102.4 \ell_c$.
First we present the results for the situation when the equilibrium diffusion coefficient $D_T(\bm{r})$ is the same inside and outside of the raft. After that we show what is changed if the equilibrium diffusion is slow within the raft.

\begin{figure}
\begin{center}
\includegraphics{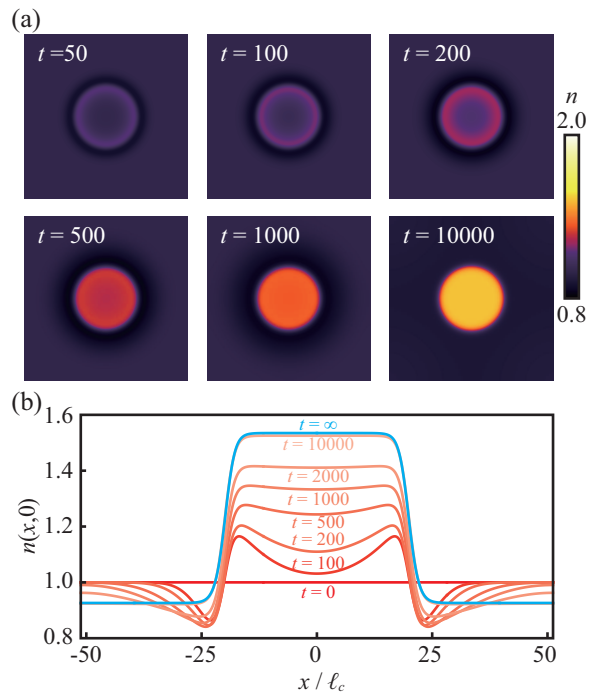}
\end{center}
\caption{(Color online) Accumulation of passive particles by a raft occupied with active proteins. Consequent snapshots (a) of the concentration distribution and radial profiles at different time moments (b) are displayed. The final profile ($t = \infty$) is determined using the analytical solution~\eqref{st_sol}. The parameters are $R = 20 \ell_c$, $\delta = 2 \ell_c$ and $\xi/D_T = 1$.}
\label{Fig_circle}
\end{figure}

\begin{figure}
\begin{center}
\includegraphics{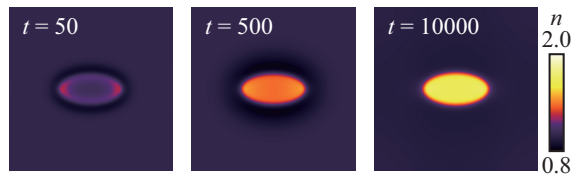}
\end{center}
\caption{(Color online) Accumulation of passive particles by an elliptic raft. Consequent snapshots of the concentration distribution at three time moments (b) are displayed. The major and minor semiaxes of the ellipse are $20 \ell_c$ and $10 \ell_c$. Other parameters are $\delta = 2 \ell_c$ and $\xi/D_T = 1$.}
\label{Fig_ellipse}
\end{figure}

\begin{figure}
\begin{center}
\includegraphics{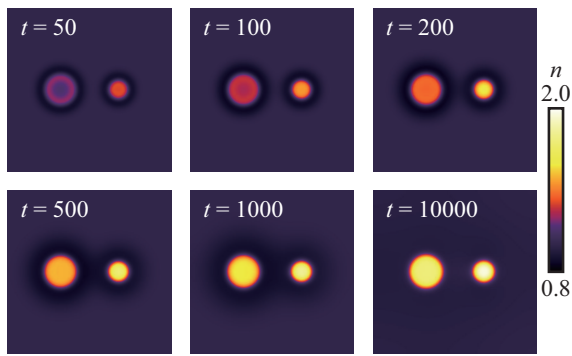}
\end{center}
\caption{(Color online) Accumulation of passive particles by two rafts of different sizes. Consequent snapshots of the concentration distribution are displayed. The parameters are $R_1 = 10 \ell_c$, $R_2 = 6 \ell_c$, $\delta = 2 \ell_c$ and $\xi/D_T = 1$.}
\label{Fig_twocircle}
\end{figure}

\begin{figure}
\begin{center}
\includegraphics{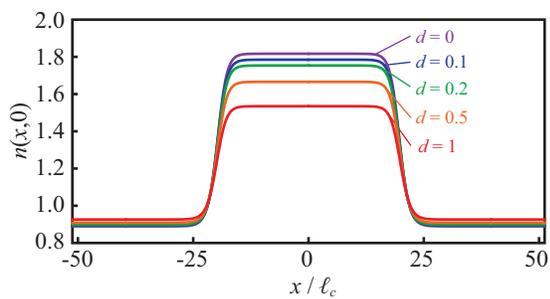}
\end{center}
\caption{(Color online) Asymptotic stationary concentration distributions of passive particles for different equilibrium diffusion coefficients inside the raft. The ratios $d = D_0 / D_\infty$ are indicated next to each curve. Other parameters are $R = 20 \ell_c$, $\delta = 2 \ell_c$ and $\xi / D_\infty = 1$.}
\label{Fig_Din}
\end{figure}

Figure~\ref{Fig_circle} shows several consequent snapshots of the concentration distribution and radial profiles of this distribution when active proteins occupy a raft of radius $R= 20 \ell_c$ and width $\delta = 2 \ell_c$, so that their distribution is given by Eq.~\eqref{distrib}. It can be seen that passive particles gradually accumulate inside the raft until a stationary distribution is formed. Note that the last profile ($t = \infty$) in Fig.~\ref{Fig_circle}b is drawn by using the analytical solution~\eqref{nin_nout}. The particles that reach by diffusion the boundary of the raft become dragged inside it and, on the other hand, the particles can also leave the raft. As a result, their concentration is depleted in the vicinity of the raft and enhanced near the boundary inside it. The depletion disappears when the final steady state is reached.
 
In Fig.~\ref{Fig_ellipse}, a similar simulation for an elliptic raft is displayed. The distribution of active proteins in this case is 
\begin{equation}
c(\bm{r}) = \frac{c_0}{2} \left[ 1 + \tanh \left( - \frac{\lambda(\mu - \mu_0)}{\delta} \right) \right]. \label{eq_ellipse}
\end{equation}
Here, elliptic coordinates $\mu$ and $\nu$ are employed and $\lambda$ is a positive value depending on $\nu$ (see Appendix~\ref{app3}). Because the raft is elongated, the accumulation of passive particles begins in the focal points, at the two ends of the ellipse. Later, however, the particles become uniformly distributed inside the raft.

Figure~\ref{Fig_twocircle} displays the results of a simulation with two rafts of different sizes. Both of them accumulate passive particles, but the accumulation proceeds faster for the smaller raft. The final concentrations of the particles in the two rafts are not much different. This could be indeed expected because of the weak logarithmic size dependence in Eq.~\eqref{ratio}. Moreover, we can see that the processes in the two rafts are roughly independent, even though the separation between the rafts is not large. 

In the above simulations, the equilibrium diffusion coefficient was assumed to be constant across the membrane. It may be however that equilibrium diffusion is slower inside the rafts. Therefore, we consider also the case when, for a single circular raft, the equilibrium diffusion coefficient depends as
\begin{equation}
D_T(\bm{r}) = D_\infty + \frac{D_0 - D_\infty}{2}\left [ 1 + \tanh \left ( -\frac{r -R}{\delta} \right ) \right ], \label{Ddistrib}
\end{equation}
on the radial coordinate, $r$. Thus, it changes from $D_0$ in the center of the raft to $D_\infty$ far from the raft. We assume that $D_0 < D_\infty$.

The equation~\eqref{st_sol} that determines the asymptotic stationary distribution of passive particles is general and it holds also if the equilibrium diffusion coefficient is coordinate-dependent. Therefore, it can be used to obtain the distributions of passive particles within a raft occupied by active proteins if the equilibrium diffusion coefficient obeys the dependence~\eqref{Ddistrib}. By numerically taking the integrals in Eq.~\eqref{st_sol}, we obtain a family of the distributions for different values of $D_0$. The results are displayed in Fig.~\ref{Fig_Din}.

It can be seen that the effect of protein activity becomes enhanced when the equilibrium diffusion is slower inside the raft. Comparing the profiles for $D_0$ / $D_\infty = 0.1$ and $1$, we see that the concentration of passive particles inside the raft is larger by about 20 percent when equilibrium diffusion in the raft is suppressed. 

It should be stressed that, when proteins inside the rafts are not active (because, for instance, ATP is not supplied), the distribution of passive particles remains uniform even if the equilibrium diffusion coefficient is decreased within the raft. Thus, the accumulation of passive particles inside is entirely due to the activity of proteins within it.

\section{Discussion}

Based on high-precision {\it in vitro} experiments~\cite{Weitz}, D.~Weitz with coworkers have recently come to the conclusion that random motion, so ubiquitous in cells, is not a result of thermally-induced fluctuations, but is instead the result of the random forces due to the aggregate motor activity in cells. Similar conclusions could be also made in other experiments~\cite{Parry}. Theoretical estimates~\cite{PNAS,Kapral} have revealed that non-thermal random forces can be the effect of nonequilibrium hydrodynamic fluctuations that are collectively induced by the activity of protein machines, including molecular motors, in the cytoplasm.

Our present analysis (see also \cite{PNAS}) suggests that a similar situation should be characteristic for biological membranes within the cells. We have shown that diffusion enhancement due to nonequilibrium hydrodynamic fluctuations in lipid bilayers can be of the same order or even stronger than diffusion due to thermal fluctuations. This means that non-thermal effects have to be taken into account when kinetic and transport phenomena within biological membranes under {\it in vivo} conditions are discussed. The intramembrane transport should strongly depend on supply of ATP (or other substrates), so that diffusion in lipid bilayers in biological cells becomes much more slow in absence of ATP. 

Because lipid bilayers behave as 2D fluids on submicrometer scales, there is however also an important difference for biomembranes as compared to the cytoplasm. Hydrodynamic 2D interactions are ultra-long ranged, depending logarithmically on the distance, and this leads to pronounced nonlocal effects. Thus, we have found that diffusion is enhanced not only within a raft occupied by active proteins, but also in the area around it. According to Eq.~\eqref{DAperp}, diffusion enhancement is proportional to the inverse square of the distance from the raft. In addition to diffusion effects, chemotaxis-like drift also takes place. However, the drift is determined only by the local distribution of active proteins.

Our analytical investigations and numerical simulations have shown that rafts occupied by active proteins tend to attract passive inclusions (such as, e.g., other membrane proteins) and accumulate them. As a result, concentration of passive particles inside the raft becomes increased. Importantly, this increase persists only as long as the proteins are active, i.e. while ATP is supplied. Thus, by varying the ATP supply to biomembranes, spatial distribution of protein inclusions can be controlled. This effect holds also if the equilibrium diffusion coefficient of passive particles inside the raft is decreased.
  
The focus in this study was on hydrodynamic effects and, to more clearly see contributions coming from them, we neglected other aspects. Thus, potential energetic interactions between proteins were not taken into account, even though they should be essential at high concentrations characteristic for the rafts. We assumed that the considered membrane is flat and limited our analysis to submicrometer length scales where coupling to the solvent is negligible. Random planar orientations of active proteins were assumed, therefore excluding the possibility of a nematic order. The most strong simplification was that fluctuations in the concentration of active proteins were not taken into account. Such fluctuations should be however relatively strong because a raft would typically include only tens of protein molecules. Therefore, the reported results should be viewed as referring to an idealized model. They are intended to demonstrate the principal hydrodynamic collective effects of active proteins in biomembranes and have to be complemented by further analytical and numerical studies. 

It should be stressed that both in Ref.~\cite{PNAS}, \cite{Kapral}, and in the present study, the situation is considered where active inclusions cyclically change their shape, but do not propel themselves, i.e., do not swim through the membrane. Formally, this corresponds to the assumption that their active shape changes are reciprocal. It has been previously noted that shape changes of enzymes within a turnover cycle can be non-reciprocal and thus self-propulsion of active proteins may take place~\cite{Sakaue}. Moreover, the propulsion effects were also demonstrated in numerical simulations for biomembranes where however model active inclusions have been used~\cite{Huang}. It is not yet clear what should be the magnitude of propulsion effects for actual protein inclusions in biological membranes and therefore we have not considered them.

In this respect, our study is different from experimental~\cite{Wu} and theoretical~\cite{Hatwalne,Underhill,Hernandez} investigations for thin fluid layers occupied by bacteria that actively change their shapes and thus swim. Diffusion can be enhanced up to a factor of 100 in such bacterial layers~\cite{Wu} and, principally, this is a similar effect. Because the bacteria swim, they cannot however form stationary rafts. Moreover, velocity correlations develop and therefore orientational nematic order in swimming bacterial populations emerge. 

It would be interesting to test the predicted effects in the experiments with actual biological membranes. In such experiments, artificially created rafts of larger sizes can be used. The activity of protein inclusions can be controlled either by varying the supply of ATP or other substrates or by chemical or optical inhibition of their turnover cycles. By repeatedly switching it on and off, flows of passive particles into the raft or out of it can then be induced. 

\begin{acknowledgments}
We are grateful for R. Kapral for stimulating discussions.
This work was supported by JSPS KAKENHI Grant Numbers JP25103008, JP26520205, and JP15K05199. This work was supported in part by the Core-to-Core Program ``Nonequilibrium dynamics of soft matter and information''(No.~23002) to Y.K. and H.K.
It was also supported by the EU program ``Non-equilibrium dynamics of soft and active matter'' (Grant No. PIRSES-GA-2011-295243) to A.M.
\end{acknowledgments}

\appendix

\section{Derivation of Eqs.~\eqref{eq5}, \eqref{eq6}, and \eqref{eq9} \label{app1}}

The diffusion tensor, $D^{A}_{\alpha {\alpha}'}$, is simplified from Eq.~\eqref{eq2} into the form which is convenient for further analysis:
\begin{align}
&{D^A}_{\alpha {\alpha}'} (\bm{r}) \nonumber \\
& = \Omega_{\beta {\beta}' \gamma {\gamma}'} \int d\bm{r}' \frac{\partial G_{\alpha \beta}}{\partial r'_\gamma} \frac{\partial G_{{\alpha}' {\beta}'}}{\partial r'_{\gamma'}} S_A (\bm{r} + \bm{r}') c(\bm{r} + \bm{r}') \nonumber \\
&= \frac 1 8 \int d\bm{r}' \left [ 
2 \left ( \frac{\partial G_{\alpha 1}}{\partial r'_1} \frac{\partial G_{{\alpha}' 1}}{\partial r'_1} + \frac{\partial G_{\alpha 2}}{\partial r'_2} \frac{\partial G_{{\alpha}' 2}}{\partial r'_2} \right ) \right. \nonumber \\ & \quad \left.
+ \left ( \frac{\partial G_{\alpha 1}}{\partial r'_2} + \frac{\partial G_{{\alpha} 2}}{\partial r'_1} \right ) 
\left ( \frac{\partial G_{{\alpha}' 1}}{\partial r'_2} + \frac{\partial G_{{\alpha}' 2}}{\partial r'_1} \right ) 
\right. \nonumber \\
& \quad \left.
+ \left ( \frac{\partial G_{\alpha 1}}{\partial r'_1} + \frac{\partial G_{\alpha 2}}{\partial r'_2} \right ) 
\left ( \frac{\partial G_{{\alpha}' 1}}{\partial r'_1} + \frac{\partial G_{{\alpha}' 2}}{\partial r'_2} \right )
\right ]  \nonumber \\ 
& \quad \times Q(\bm{r} + \bm{r}') \nonumber \\
&= \frac{1}{32 \pi^2 \eta^2} \int d\bm{r}' \frac{r'_\alpha r'_{{\alpha}'}}{{r'}^4} Q(\bm{r} + \bm{r}'),
\end{align}
where $Q(\bm{r}) = S_A(\bm{r}) c(\bm{r})$. Here, we use 
\begin{align}
\Omega_{\beta {\beta}' \gamma {\gamma}'} = \left \{
\begin{array}{ll}
\displaystyle{\frac 3 8}, & \mathrm{if} \; (\beta, {\beta}', \gamma, {\gamma}') = (1,1,1,1), (2,2,2,2), \\
\displaystyle{\frac 1 8}, & \mathrm{if} \;(\beta, {\beta}', \gamma, {\gamma}') = (1,1,2,2), (1,2,1,2), \\
& (1,2,2,1), (2,1,1,2), (2,1,2,1), (2,2,1,1), \\
0, & \mathrm{otherwise},
\end{array}
\right.
\end{align}
and the first and second derivatives of the Oseen tensor in Eq.~\eqref{2DOseen},
\begin{equation}
\frac{\partial G_{\alpha \beta}}{\partial r_\gamma} = \frac{1}{4\pi\eta} \left \{ \frac{1}{r^2} (-r_\gamma \delta_{\alpha \beta} + r_\alpha \delta_{\beta \gamma} + r_\beta \delta_{\alpha \gamma}) - \frac{2r_\alpha r_\beta r_\gamma}{r^4} \right \},
\end{equation}
\begin{align}
\frac{\partial^2 G_{\alpha \beta}}{\partial r_\gamma \partial r_\delta} 
=& \frac{1}{4\pi\eta} \left \{ \frac{1}{r^2} (-\delta_{\alpha \beta} \delta_{\gamma \delta} + \delta_{\alpha \delta} \delta_{\beta \gamma} + \delta_{\alpha \gamma} \delta_{\beta \delta}) \right. \nonumber \\
& - \left. \frac{2}{r^4} (-r_\gamma r_\delta \delta_{\alpha \beta} + r_\alpha r_\beta \delta_{\gamma \delta} + r_\alpha r_\gamma \delta_{\beta \delta}  \right. \nonumber \\
 & \left. + r_\alpha r_\delta \delta_{\beta \gamma} + r_\beta r_\gamma \delta_{\alpha \delta} + r_\beta r_\delta \delta_{\alpha \gamma} ) + \frac{8r_\alpha r_\beta r_\gamma r_\delta}{r^6} \right \}.
\end{align}

The velocity, $V_\alpha (\bm{r})$, is also simplified from Eq.~\eqref{eq3} in the same way.
\begin{align}
&V_\alpha (\bm{r}) \nonumber \\
&= -\Omega_{\beta {\beta}' \gamma {\gamma}'} \int d\bm{r}' \frac{\partial^2 G_{\alpha \beta}}{\partial r'_\gamma \partial r'_\delta} \frac{\partial G_{\delta {\beta}'}}{\partial r'_{\gamma'}} S_A (\bm{r} + \bm{r}') c(\bm{r} + \bm{r}') \nonumber \\
&= - \frac{1}{8} \int d\bm{r}' \left [ 
2 \left ( \frac{\partial^2 G_{\alpha 1}}{\partial r'_1 \partial r'_\delta} \frac{\partial G_{\delta 1}}{\partial r'_1} 
+ \frac{\partial^2 G_{\alpha 2}}{\partial r'_2 \partial r'_\delta} \frac{\partial G_{\delta 2}}{\partial r'_2} \right ) \right. \nonumber \\
& \quad \left. + \left ( \frac{\partial^2 G_{\alpha 1}}{\partial r'_2 \partial r'_\delta} + \frac{\partial^2 G_{\alpha 2}}{\partial r'_1 \partial r'_\delta} \right ) 
\left ( \frac{\partial G_{\delta 1}}{\partial r'_2} + \frac{\partial G_{\delta 2}}{\partial r'_1} \right ) 
\right. \nonumber \\
& \quad \left.
+ \left ( \frac{\partial^2 G_{\alpha 1}}{\partial r'_1 \partial r'_\delta} + \frac{\partial^2 G_{\alpha 2}}{\partial r_2 \partial r_\delta} \right ) 
\left ( \frac{\partial G_{\delta 1}}{\partial r'_1} + \frac{\partial G_{\delta 2}}{\partial r'_2} \right )
\right ] \nonumber \\
& \quad \times Q(\bm{r} + \bm{r}') \nonumber \\
&= \frac{1}{32 \pi^2 \eta^2} \int d\bm{r}' \frac{r'_\alpha}{{r'}^4} Q(\bm{r} + \bm{r}').
\end{align}

Finally, the derivation for Eq.~\eqref{eq9} is shown. From Eq.~\eqref{eq8}, we calculate
\begin{align}
U_\alpha (\bm{r}) 
=& V_\alpha (\bm{r}) - \frac{\partial D_{\alpha \alpha'} (\bm{r})}{\partial r_{\alpha'}} \nonumber \\
=& \frac{1}{32 \pi^2 \eta^2} \int d\bm{r}' \left ( \frac{r'_\alpha}{{r'}^4} -  \frac{r'_\alpha r'_{\alpha'}}{{r'}^4} \frac{\partial}{\partial r_{\alpha'}}\right ) Q (\bm{r} + \bm{r}') \nonumber \\
=& \frac{1}{32 \pi^2 \eta^2} \int d\bm{r}' \left ( \frac{r'_\alpha}{{r'}^4} + \left( \frac{\partial}{\partial r'_{\alpha'}} \frac{r'_\alpha r'_{\alpha'}}{r'^4} \right) \right ) Q (\bm{r} + \bm{r}') \nonumber \\  &- \frac{1}{32 \pi^2 \eta^2} \int_{\sigma} d s'_{\alpha'} \frac{r'_\alpha r'_{\alpha'}}{{r'}^4} Q (\bm{r} + \bm{r}'),
\end{align}
where $\int_{\sigma} d s'_{\alpha'}$ is the integration along the periphery of the domain.  Here, $\partial/\partial r_{\alpha'}$ can be regarded as $\partial/\partial r'_{\alpha'}$ and the partial integration is used. 
The derivative in the integrand is calculated as
\begin{align}
\frac{\partial}{\partial r_{\alpha'}} \frac{r_\alpha r_{\alpha'}}{r^4} 
= \frac{\delta_{\alpha \alpha'} r_{\alpha'}}{r^4} + 2 \frac{r_\alpha}{r^4} - 4 \frac{r_\alpha r_{\alpha'}^2}{r^6} = - \frac{r_\alpha}{r^4}.
\end{align}
Thus, only the surface term remains
\begin{align}
U_\alpha (\bm{r}) 
= - \frac{1}{32 \pi^2 \eta^2} \int_{\sigma} d s'_{\alpha'} \frac{r'_\alpha r'_{\alpha'}}{{r'}^4} Q (\bm{r} + \bm{r}'). \label{dif_U}
\end{align}
The integration is taken over the physical boundary $\sigma_{\mathrm{outside}}$ and the small cut-off surface $\sigma_{\mathrm{inside}}$ around $\bm{r}$. The integration taken over the physical boundary $\sigma_{\mathrm{outside}}$ becomes zero if $Q = 0$ at the boundary, as we always assume.
As for the cut-off surface, we expand $Q$ as
\begin{equation}
Q (\bm{r} + \bm{r}') = Q(\bm{r}) + r'_{\alpha} \frac{\partial  Q(\bm{r})}{\partial r_{\alpha}} + \mathcal{O}({r'}^2).
\end{equation}
Then, the integral over the small cut-off surface is calculated as
\begin{align}
U_\alpha (\bm{r}) 
=& - \frac{1}{32 \pi^2 \eta^2} \int_{0}^{2\pi} (-\ell_c \hat{r'}_{\alpha'} d \phi') \frac{\hat{r'}_\alpha \hat{r'}_{\alpha'}}{{\ell_c}^2} \nonumber \\
& \times \left( Q(\bm{r}) + \ell_c \hat{r'}_{\beta} \frac{\partial Q(\bm{r})}{\partial r_{\beta}} + \mathcal{O}({\ell_c}^2) \right ) \nonumber \\
=& \frac{1}{32 \pi \eta^2} \frac{\partial Q(\bm{r})}{\partial r_{\alpha}} + \mathcal{O}(\ell_c),
\end{align}
where $\hat{r'}_{\alpha}$ is a unit vector which is parallel to $r'_{\alpha}$, and $\hat{r'}_1 = \cos \phi'$ and $\hat{r'}_2 = \sin \phi'$. Here, we used $\hat{r'}_\alpha \hat{r'}_{\alpha} = 1$, and the integrations of $\hat{r}_\alpha$ and $\hat{r}_\alpha \hat{r}_{\alpha'}$ with regard to $\phi$ over $[0,2\pi)$ are $0$ and $\pi \delta_{\alpha \alpha'}$, respectively.

\section{Analytical solution for a circular raft\label{app2}}

In this section, the drift velocity and diffusion enhancement are derived for a circular raft with a radius of $R$.
First, the case when the passive particle is located inside the raft, i.e., $0 \leq r < R-\ell_c$, is considered.
We calculate the drift velocity $V_{\alpha}$ and diffusion enhancement $D^A_{\alpha \alpha'}$ at the position $\bm{r}=(r,0)$ in the Cartesian coordinates. 
We adopt the polar coordinates whose origin corresponds to $\bm{r}$.
The range of the integral in the radial direction is $\displaystyle{\left [\ell_c, -r \cos \theta + \sqrt{R^2 - r^2 \sin^2 \theta} \right ]}$. 
The lower limit of the integral is the cut-off length and the upper limit is obtained by solving ${r_{\mathrm{max}}}^2 + r^2 - 2 r_{\mathrm{max}} r \cos (\pi - \theta) = R^2$ with regard to $r_{\mathrm{max}}$.
Here, we define $L(\theta)$ as $L(\theta) = -r \cos \theta + \sqrt{R^2 - r^2 \sin^2 \theta}$.
The drift velocity is derived as
\begin{align}
\bm{V} (\bm{r}) =& \frac{S_A}{32 \pi^2 \eta^2} \int d\bm{r}' \frac{1}{{r'}^4} \left (
\begin{array}{c}
r'_1 \\
r'_2 
\end{array}
\right ) c(\bm{r}+\bm{r}') \nonumber \\
=& \xi \int_{0}^{2\pi} d\theta \int_{\ell_c}^{L(\theta)} r'dr' \frac{1}{{r'}^4} \left (
\begin{array}{c}
r' \cos \theta \\
r' \sin \theta
\end{array}
\right ) \nonumber \\
=& \pi \xi \left (
\begin{array}{c}
\displaystyle{-\frac{r}{R^2-r^2}} \\
0
\end{array}
\right ),
\end{align}
where $c(\bm{r})$ is defined as 
\begin{align}
c(\bm{r}) = \left \{
\begin{array}{ll}
c_0, & \mathrm{if} \; |\bm{r}|<R, \\
0, & \mathrm{if} \; |\bm{r}|>R,
\end{array}
\right .
\end{align}
and $\xi = S_A c_0 / (32 \pi^2 \eta^2)$. 
The diffusion enhancement is derived as
\begin{align}
D^A (\bm{r}) 
=& \frac{S_A}{32 \pi^2 \eta^2} \int d\bm{r}' \frac{1}{{r'}^4} \left (
\begin{array}{cc}
{r'_1}^2 & r'_1 r'_2 \\
r'_1 r'_2 & {r'_2}^2
\end{array}
\right ) c(\bm{r}+\bm{r}') \nonumber \\
=& \xi \int_{0}^{2\pi} d\theta \int_{\ell_c}^{L(\theta)} r'dr' \frac{1}{{r'}^4} \nonumber \\
&\times \left (
\begin{array}{cc}
{r'}^2 \cos^2 \theta & {r'}^2 \sin \theta \cos \theta \\
{r'}^2 \sin \theta \cos \theta & {r'}^2 \sin^2 \theta
\end{array}
\right ) \nonumber \\
=& \pi \xi \left ( \ln \frac{\sqrt{R^2 - r^2}}{\ell_c} \right ) I,
\end{align}
where $I_{\alpha \beta} = \delta_{\alpha \beta}$. Thus, Eq.~\eqref{diffin} is obtained.

Second, the case when the passive particle is located outside of the raft, i.e., $r > R + \ell_c$, is considered.
The drift velocity is calculated as
\begin{align}
\bm{V} (\bm{r}) 
=& \frac{S_A}{32 \pi^2 \eta^2} \int d\bm{r}' \frac{1}{{r'}^4} \left (
\begin{array}{c}
r'_1 \\
r'_2 
\end{array}
\right ) c(\bm{r}+\bm{r}') \nonumber \\
=& \xi \int_{0}^{R} r'dr' \int_{0}^{2\pi} d\theta \frac{1}{({r'}^2 + r^2 - 2 r' r \cos \theta)^2} \nonumber \\
& \times \left (
\begin{array}{c}
r' \cos \theta - r\\
r' \sin \theta
\end{array}
\right ) \nonumber \\
=& \pi \xi \left (
\begin{array}{c}
\displaystyle{-\frac{R^2}{r(r^2-R^2)}} \\
0
\end{array}
\right ).
\end{align}
The diffusion enhancement is derived as
\begin{align}
D^A (\bm{r}) 
=& \frac{S_A}{32 \pi^2 \eta^2} \int d\bm{r}' \frac{1}{{r'}^4} \left (
\begin{array}{cc}
{r'_1}^2 & r'_1 r'_2 \\
r'_1 r'_2 & {r'_2}^2
\end{array}
\right ) c(\bm{r}+\bm{r}') \nonumber \\
=& \xi \int_{0}^{R} r'dr' \int_{0}^{2\pi} d\theta \frac{1}{({r'}^2 + r^2 - 2 r' r \cos \theta)^2}
\nonumber \\
& \times
 \left (
\begin{array}{cc}
(r' \cos \theta - r)^2 & (r' \cos \theta - r) r' \sin \theta \\
r' \sin \theta (r' \cos \theta - r) & {r'}^2 \sin^2 \theta
\end{array}
\right ) \nonumber \\
=& \pi \xi 
\left (
\begin{array}{cc}
\displaystyle{\ln \frac{r}{\sqrt{r^2 - R^2}} + \frac{R^2}{2 r^2}} & 0 \\
0 & \displaystyle{\ln \frac{r}{\sqrt{r^2 - R^2}} -\frac{R^2}{2 r^2}}
\end{array}
\right ).
\end{align}
Thus, Eqs.~\eqref{diffinn} and \eqref{diffout} are obtained.

In addition, the analytical solution near the periphery of the raft, i.e. $R-\ell_c < r < R  + \ell_c$, is obtained only for the radial component of the velocity, $V_{\|}$ as follows:
For $R-\ell_c \leq r < \sqrt{R^2-{\ell_c}^2}$,
\begin{widetext}
\begin{align}
V_{\|}(r) =& \int_{\theta_0}^{2\pi-\theta_0} d\theta \int_{\ell_c}^{a(\theta)} rdr \frac{1}{r^4} r \cos \theta \nonumber \\
=& \xi \left [
\frac{r}{R^2-r^2} \left (\frac{R^2}{r^2} \arccos \frac{R^2+{\ell_c}^2-r^2}{2 R \ell_c} + \arccos \frac{R^2-r^2-{\ell_c}^2}{2 r \ell_c} -\pi \right )  - \frac{\sqrt{(R^2-(r-\ell_c)^2)((r+\ell_c)^2-R^2)}}{2 r {\ell_c}^2} \right ],
\end{align}
\end{widetext}
where $a(\theta) = - r \cos \theta + \sqrt{R^2 - r^2 \sin^2 \theta}$ and $\theta_0 = \arccos [(R^2-r^2-{\ell_c}^2)/(2 r \ell_c)] < \pi/2$.

For $\sqrt{R^2-{\ell_c}^2} < r < \sqrt{R^2+{\ell_c}^2}$,
\begin{widetext}
\begin{align}
V_{\|}(r) =& \int_{\theta_0}^{2\pi-\theta_0} d\theta \int_{\ell_c}^{a(\theta)} rdr \frac{1}{r^4} r \cos \theta \nonumber \\
=& \xi \left [
\frac{r}{R^2-r^2} \left ( \frac{R^2}{r^2} \arccos \frac{R^2+{\ell_c}^2-r^2}{2 R \ell_c}  -\arccos \frac{r^2 -R^2 +{\ell_c}^2}{2 r \ell_c} \right ) - \frac{\sqrt{(R^2-(r-\ell_c)^2)((r+\ell_c)^2-R^2)}}{2 r {\ell_c}^2} \right ],
\end{align}
\end{widetext}
where $a(\theta) = - r \cos \theta + \sqrt{R^2 - r^2 \sin^2 \theta}$ and $\theta_0 = \arccos [(R^2-r^2-{\ell_c}^2)/(2 r \ell_c)] > \pi/2$.

\begin{figure}
	\begin{center}
		\includegraphics{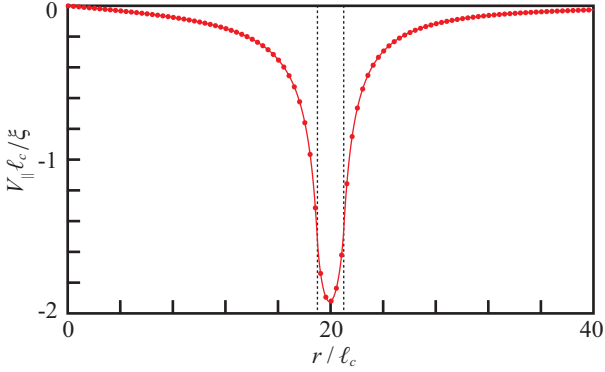}
		\caption{Profiles of the radial component of the velocity, $V_{\|}$ obtained by the numerical integration (closed circles) and analytical calculation (solid curves) in the case with a circular raft of a radius $R$, in which the concentration of active proteins is $c_0$. We set $R/\ell_c = 20$, just as in Fig.~\ref{Fig_profile_delta1}. The parameter $\delta$ in Eq.~\eqref{distrib} is set as $\delta=0$. 
}
		\label{fig-app-vx}
	\end{center}
\end{figure}

For $\sqrt{R^2+{\ell_c}^2} < r < R+\ell_c$,
\begin{widetext}
\begin{align}
V_{\|}(r) =& \int_{\theta_0}^{2\pi-\theta_0} d\theta \int_{\ell_c}^{a(\theta)} rdr \frac{1}{r^4} r \cos \theta \nonumber \\
=& \xi \left [
-\frac{\pi R^2}{r (r^2-R^2)} + \frac{r}{R^2-r^2} \left ( -\arccos \frac{r^2 +{\ell_c}^2 -R^2}{2 r \ell_c} -\frac{R^2}{r^2} \arccos \frac{r^2-R^2-{\ell_c}^2}{2 R \ell_c} \right ) \right . \nonumber \\
& \left . \quad \quad \quad \quad -\frac{\sqrt{(R^2-(r-\ell_c)^2)((r+\ell_c)^2-R^2)}}{2 r {\ell_c}^2} \right ],
\end{align}
\end{widetext}
where $a(\theta) = - r \cos \theta - \sqrt{R^2 - r^2 \sin^2 \theta}$ and $\theta_0 = \arccos [(R^2-r^2-{\ell_c}^2)/(2 r \ell_c)] > \pi/2$.
It is noted that the transverse component of the velocity, $V_{\bot}$, is always zero due to the symmetry of the system.
The analytical results are shown in Fig.~\ref{fig-app-vx} together with the numerical results. The analytical solution well matches the numerical results.

We have also obtained the explicit form of $\bm{U}(\bm{r})$ analytically.
From the expression \eqref{dif_U}, it is easily concluded that $\bm{U}=\bm{0}$ for $r<R-\ell_c$ and $R+\ell_c < r$, and $U_{\bot}=0$ for all region.
Thus, only $U_{\|}$ for $R-\ell_c < r < R+\ell_c$ is non-trivial and is calculated as 
\begin{equation}
U_{\|}(r) = -\xi \frac{\sqrt{\left( R^2 - (r - \ell_c)^2 \right) \left( (r + \ell_c)^2 - R^2 \right)}}{r {\ell_c}^2}.
\end{equation}
When $\ell_c$ is sufficiently small, $U_1(r)$ is approximated as
\begin{equation}
U_{\|}(r) = -\pi \xi \delta(r - R),
\end{equation}
and this result appears in Eq.~\eqref{delta_fn}.

\section{Smoothed profile of an elliptic raft\label{app3}}

The elliptic coordinates $\mu$ and $\nu$ are defined as
\begin{equation}
x = \chi \cosh \mu \cos \nu,
\end{equation}
\begin{equation}
y = \chi \sinh \mu \sin \nu,
\end{equation}
where $\chi$ is a positive parameter. The level curve of $\mu = \mu_0 > 0$ corresponds to an ellipse, 
\begin{equation}
\frac{x^2}{a^2} + \frac{y^2}{b^2} = 1,
\end{equation}
where $a = \chi \cosh \mu_0$, $b = \chi \sinh \mu_0$, and $a > b > 0$. Here, $a$ and $b$ are major and minor semiaxes, respectively. 
Thus, $\chi$ is calculated as
\begin{equation}
\chi = \sqrt{a^2 - b^2}.
\end{equation}
Here it is noted that the periphery is defined as
\begin{equation}
\mu = \mu_0 = {\rm arctanh} \frac{b}{a}.
\end{equation}
Considering the level curves of $\mu$ and $\nu$ cross perpendicularly at every point, we can describe the profile as a function of only $\mu$ as
\begin{equation}
c(\bm{r}) = \frac{c_0}{2} \left[ 1 + \tanh \left( - \frac{(\mu - \mu_0)}{\delta_\mu} \right) \right],
\end{equation}
where $\delta_\mu$ is the smoothing factor. $\delta_\mu$ should be determined so that it corresponds to $\delta$ in the Cartesian coordinates. The length unit $d\ell$ along the level curve of $\nu$ is described as
\begin{equation}
\frac{d\ell}{d\mu} = \chi \sqrt{\sinh^2 \mu + \sin^2 \nu}.
\end{equation}
Therefore, $\delta_\mu$ is described as
\begin{equation}
\delta_\mu = \frac{\delta}{\lambda},
\end{equation}
where
\begin{equation}
\lambda = \chi \sqrt{\sinh^2 \mu_0 + \sin^2 \nu},
\end{equation}
and thus Eq.~\eqref{eq_ellipse} is obtained.

\end{document}